\documentclass[twocolumn,showpacs,preprintnumbers,amsmath,amssymb,prl,superscriptaddress]{revtex4}
\usepackage{graphicx}
\usepackage{epstopdf}
\usepackage{dcolumn}
\usepackage{bm}
\usepackage{amsmath}
\usepackage{amssymb}
\usepackage{mathrsfs}
\usepackage{amsfonts}
\usepackage{color}

\begin{document} 

\title{Steady-state phases and tunneling-induced instabilities \\ in the driven-dissipative Bose-Hubbard model}

\author{Alexandre Le Boit\'e}
\author{Giuliano Orso}
\author{Cristiano Ciuti}
\affiliation{Laboratoire Mat\'eriaux et Ph\'enom\`enes Quantiques,
Universit\'e Paris Diderot-Paris 7 and CNRS, \\ B\^atiment Condorcet, 10 rue
Alice Domon et L\'eonie Duquet, 75205 Paris Cedex 13, France}

\begin{abstract}

We determine the steady-state phases of a driven-dissipative Bose-Hubbard model, describing, e.g., an array of coherently pumped nonlinear cavities with a finite photon lifetime.
Within a mean-field master equation approach using exact quantum solutions for the one-site problem, we show that the system exhibits a tunneling-induced transition between monostable and bistable phases.
We characterize the corresponding quantum correlations, highlighting the essential differences with respect to the equilibrium case.
We also find collective excitations with a flat energy-momentum dispersion over the entire Brillouin zone that trigger modulational instabilities at specific wavevectors.

\end{abstract}
\pacs{42.50.Ar,03.75.Lm,42.50.Pq,71.36.+c}

\maketitle

In recent years, the interest in the physics of quantum fluids of light in systems with effective photon-photon interactions has triggered many exciting investigations \cite{RMP}. Some of the most remarkable features of quantum fluids,  such as superfluid propagation \cite{Amo09,Tanese2012} or generation of topological excitations \cite{Pigeon, Amo11, Nardin,Sanvitto} have been observed in experiments with solid-state microcavities. With the dramatic experimental advances in solid-state cavity and circuit quantum electrodynamics (QED), a considerable interest is growing on the physics of controlled arrays of nonlinear cavity resonators, which can be now explored in state-of-art systems \cite{Houck,Deveaud}.
This opens the way to the implementation of 
non-equilibrium lattice models of interacting bosons, particularly when effective on-site  photon-photon interactions are large enough to enter the strongly correlated regime\cite{Imamoglu, Fink, Liew, Bamba}. In this kind of systems, it is possible to realize the celebrated Bose-Hubbard model \cite{Fisher} for photons or polaritons. Since the first theoretical proposal for implementing this model in optical systems \cite{Hartmann06, Greentree, Angelakis}, early works have been focused on phenomena close to the equilibrium Mott insulator-Superfluid quantum phase transition\cite{Tomadin, Wu}.  Strongly-non equilibrium effects have been addressed only more recently \cite{Carusotto09, Hartmann10, Nissen,Carusotto2012} particularly in the interesting driven-dissipative regime where the cavity resonators are excited by a coherent pump which competes with the cavity dissipation processes.  
In such non-equilibrium conditions, these open systems are driven into steady-state phases whose collective excitations
can be extremely different from the equilibrium case. However, to the best of our knowledge, very little is known so far on these important properties for the non-equilibrium Bose-Hubbard model.

In this Letter, we present comprehensive results for the steady-state phases and excitations of the driven-dissipative Bose-Hubbard model in the case of homogeneous coherent pumping. The steady-state density matrix and expectation values of the relevant observables have been calculated with an efficient mean-field approach, based on exact analytical quantum optical solutions of the single-cavity problem.
A rich diagram is shown with multiple steady-state phases, whose stability and complex energy excitation spectrum have been studied through a linearization of the Lindblad master equation around the stationary solutions. We unveil the existence of a purely imaginary excitation branch which can trigger modulational instabilities at specific wavevectors.

\begin{figure}
\includegraphics[width=\columnwidth]{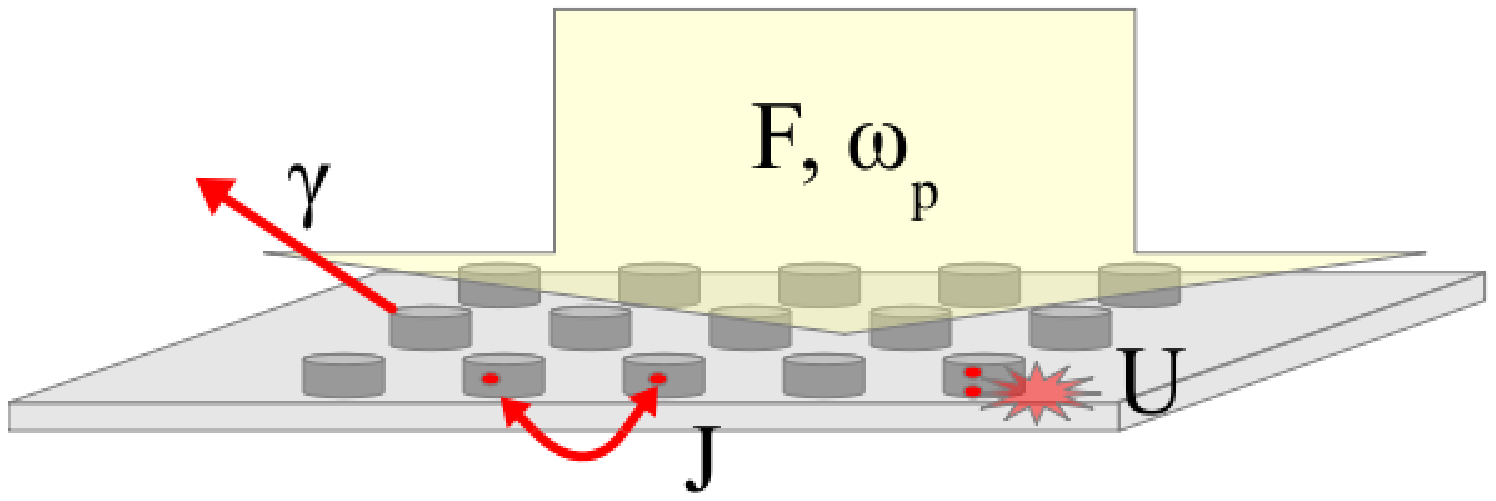}\\
\includegraphics[width = \columnwidth]{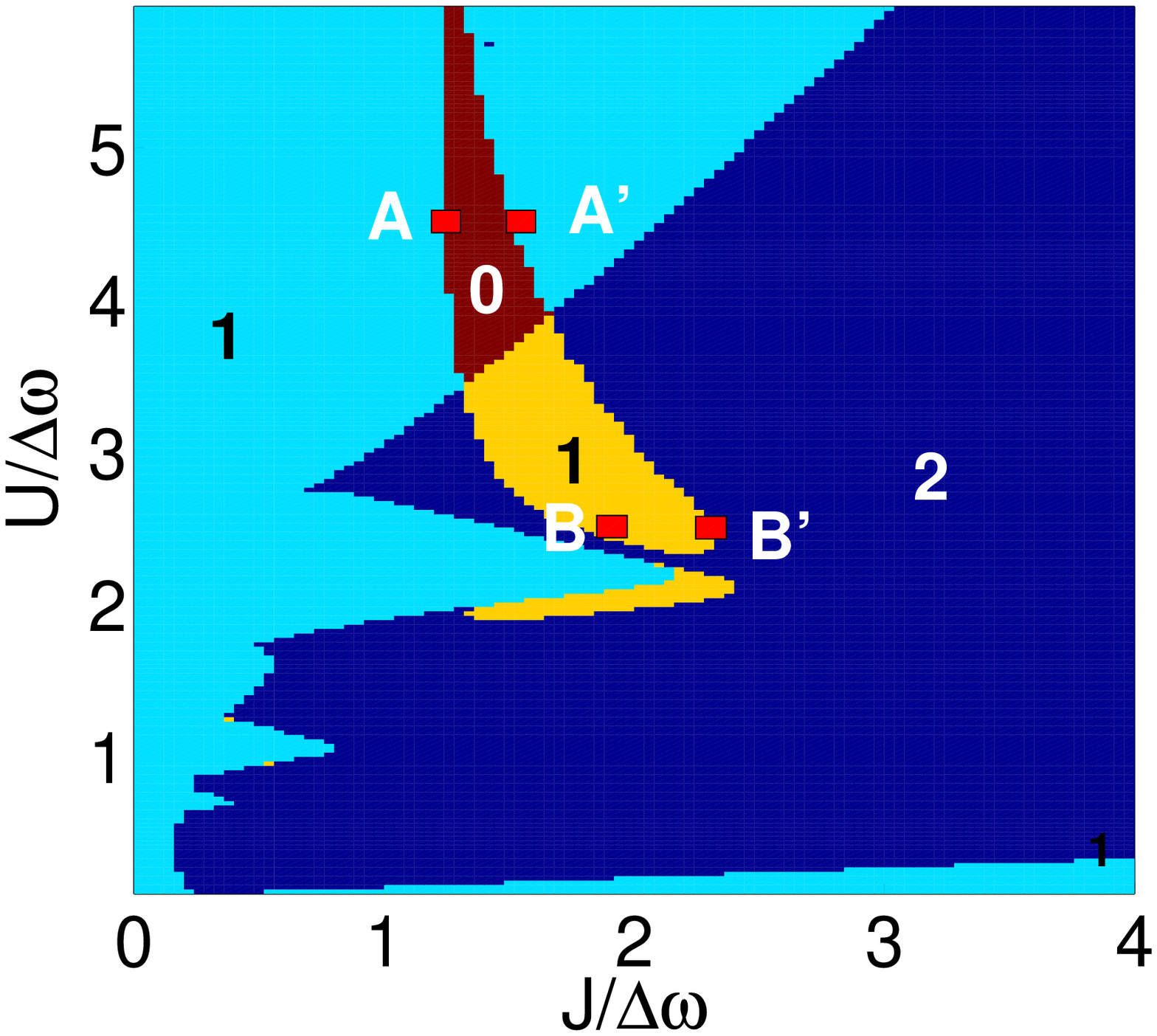}
	\caption{Color online. Top: Sketch of a square photonic lattice made of nonlinear cavities coupled by tunneling. The system is pumped coherently by an homogeneous laser field at normal incidence.  Bottom: Number of mean-field solutions and their stability plotted as a function of $J/\Delta \omega$ and  $U/\Delta \omega$, for $F/\Delta \omega = 0.4$ ; $\gamma /\Delta \omega = 0.2$ and $\Delta \omega > 0$.  Light blue (top-left and bottom-right part labeled with a `1'): monostable region, only one solution  to Eq.(\ref{self-consistent}). Dark blue part (label `2'): bistable region, two solutions to Eq.(\ref{self-consistent}). Yellow part (central region labeled with `1') has only one stable phase out of  two existing solutions. Red part (label `0'): only one solution, which is unstable.} 
\label{fig:bistab}

\end{figure}

We consider a driven-dissipative Bose-Hubbard model under homogeneous coherent pumping describing a bidimensional square lattice of cavity resonators. In a frame rotating at the pump frequency $\omega_p$, the system is described by the following Hamiltonian \cite{RMP}:
\begin{equation}
H =  -\frac{J}{z}\sum_{<i,j>}b^{\dagger}_i b_j-\sum_i^{N}\Delta \omega b^{\dagger}_i b_i+\frac{U}{2}b^{\dagger}_i b^{\dagger}_i b_ib_i + Fb^{\dagger}_{i}+ F^{*}b_{i}
\end{equation}
 where $b^{\dagger}_i$ creates a boson on site $i$, $J > 0$ is the tunneling strength,  and $z=4$ is the coordination number. $<i,j>$ indicates that tunneling is possible only between first-neighbors. $U > 0$ represents the effective on-site repulsion, $F$ is the amplitude of the incident laser field and $\Delta \omega=\omega_p-\omega_c$  is the frequency detuning of the pump with respect to the cavity mode. The dynamics of the many-body density matrix $\rho(t)$ is described in terms of the Lindblad master equation:
\begin{equation}
i\partial_{t} \rho = [H, \rho] + \frac{i\gamma}{2}\sum_{i}^{N} 2b_{i} \rho b_{i}^{\dagger} - b_{i}^{\dagger} b_{i} \rho - \rho  b_{i}^{\dagger} b_{i} \label{ME}
\end{equation}
where $\gamma$ is the dissipation rate. While for equilibrium quantum gases, the chemical potential $\mu$ is a key quantity, in this non-equilibrium model the steady-state phases depend instead on the pump parameters
$F$ and $\Delta \omega$, which compete with  $\gamma$.
  
Given the success of mean-field theories in the investigation of the equilibrium Bose-Hubbard physics, it is a legitimate starting point for the study of its non-equilibrium version. The mean-field approximation is obtained by replacing $b^{\dagger}_{i}b_{j}$ with $\langle b^{\dagger}_i \rangle b_j+\langle b_j \rangle b^{\dagger}_i$ in the many-body hamiltonian. The initial problem is then reduced to a single-site Hamiltonian describing an isolated cavity with effective pumping term $F-J\langle b\rangle$ :
\begin{equation}
H_{mf} = -\Delta \omega b^{\dagger} b+\frac{U}{2}b^{\dagger} b^{\dagger} bb + (F-J\langle b \rangle)b^{\dagger}+ (F^{*}-J \langle b \rangle ^{*})b
\end{equation}
where the value of $ \langle b \rangle $ has to be determined self-consistently. 
The problem of a single cavity has been studied by Drummond and Walls \cite{DW}, who obtained analytical expressions for the bosonic coherence $ \langle b \rangle $ and the photons distribution functions via a generalized $P$-representation for the density matrix. By replacing $F$ with $F-J\langle b \rangle$ in these exact expressions we find the self-consistent formula

\begin{equation}
\langle b \rangle = \frac{(F-J\langle b \rangle)}{\Delta \omega+i\gamma/2}\times\frac{\mathcal{F}(1+c,c^{*},8|\frac{F-J\langle b \rangle}{U}|^2)}{\mathcal{F}(c,c^{*},8|\frac{F-J\langle b \rangle}{U}|^2)} \label{self-consistent}
\end{equation}
for the bosonic coherence. The mean photon density and the other diagonal correlation functions
 can then be easily extracted from
the  general expression: 
\begin{align}
\langle (b^{\dagger})^j(b)^j \rangle =& \left|\frac{2(F-J\langle b \rangle)}{U}\right|^{2j}\times\frac{\Gamma(c)\Gamma(c^*)}{\Gamma(c+j)\Gamma(c^*+j)} \nonumber \\ 
&\times\frac{\mathcal{F}(j+c,j+c^{*},8|F/U|^2)}{\mathcal{F}(c,c^{*},8|F/U|^2)} \label{expect}
\end{align}
with
$
c = 2(-\Delta \omega -i\gamma/2)/U
$
and the hypergeometric function $
\mathcal{F}(c,d,z) = \sum_n^{\infty} \frac{\Gamma (c) \Gamma (d)}{\Gamma(c+n) \Gamma(d+n)}\frac{z^{n}}{n!}
$,
 $\Gamma$ being the gamma special function.

All the properties of the steady-states are therefore determined by the self-consistent solutions of Eq.(\ref{self-consistent}) which we have calculated numerically.
Due to the presence of the tunneling term $J$, multiple solutions do appear in certain region of parameters space. 
We investigate their stability  through a linearization of the Lindblad master equation around each steady-state solution as described later in the Letter. 
In Fig. \ref{fig:bistab}, we present a diagram showing the number of stable steady-state solutions
as a function of the tunneling and the on-site interaction in units of the detuning $\Delta\omega > 0$ and for a representative set of parameters (see caption).  We see that there are regions  with $1$ or $2$ stable solutions, but also regions
with no stable homogeneous solution (shown in red). 
Notice however that, within our mean field approach,  we have direct access only to spatially uniform solutions, where all the cavity sites are equivalent. 
  
\begin{figure}[t!]
\begin{center}
	\includegraphics[width = 0.49\columnwidth]{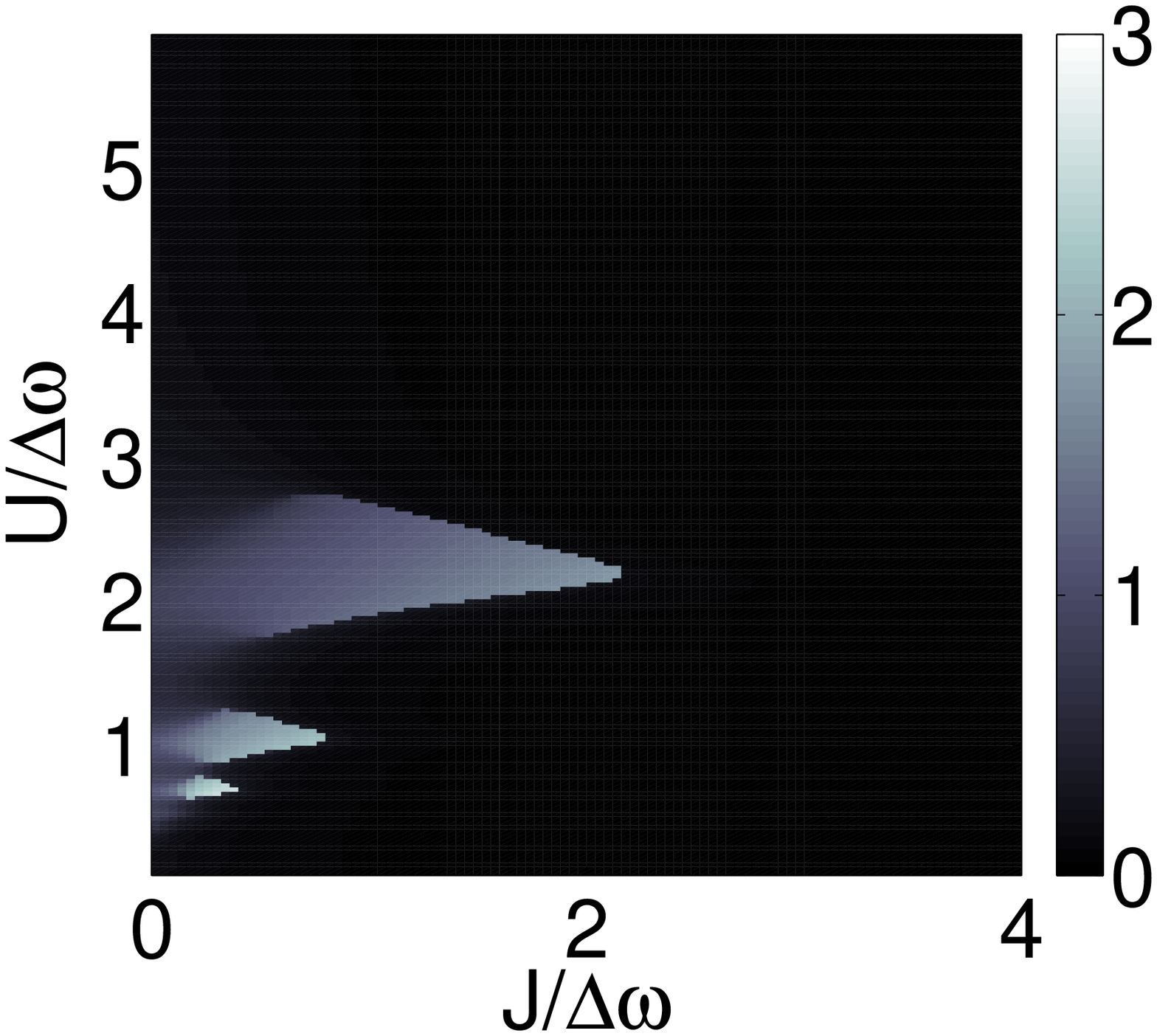}
	\includegraphics[width = 0.49\columnwidth]{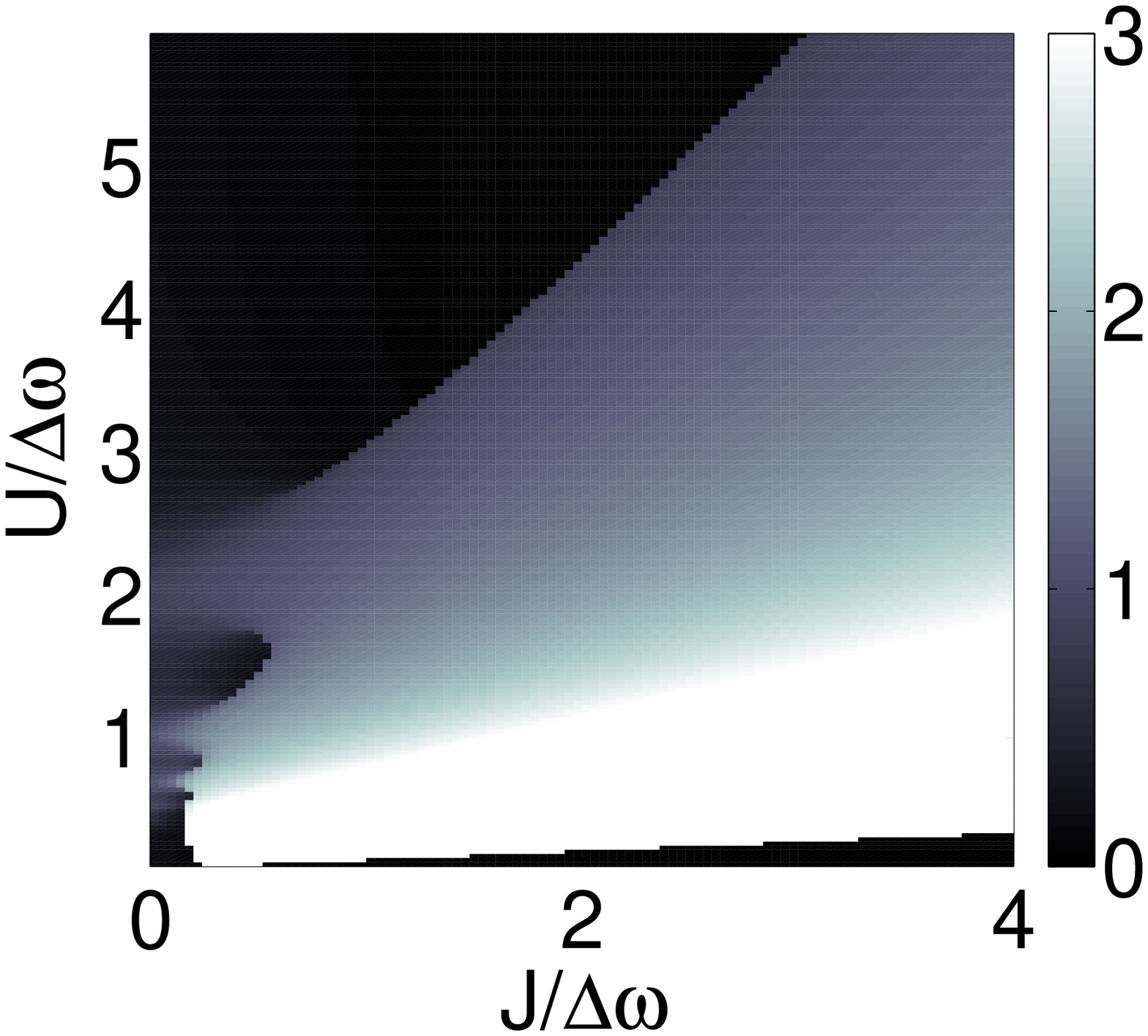}
\caption{ Photon occupation number as a function of $J/\Delta \omega$ and  $U/\Delta \omega$ for $F/\Delta \omega = 0.4$ and $\gamma /\Delta \omega = 0.2$. Left panel: results for the boson occupation number in the low-density phase. Right panel: the same quantity but for the high-density phase.  For sake of clarity, the maximal value of the colorscale in the high-density phase has been set to $3$, but the density is higher than 10 at high $J$ and low $U$. Notice that the monostable region is the same for both  panels.}
\label{fig:dens}
\end{center}
\end{figure}

Interestingly, we find that the bistability induced by the coupling between the cavities also appears when the pump frequency is red-detuned with respect to the cavity mode (not shown), in sharp contrast with the case of an isolated cavity.
 We see in Fig. \ref{fig:bistab} that the boundary between monostable and bistable phases is reminiscent of the lobe structure characteristic of 
the equilibrium model. But despite similar shapes, equilibrium and non-equilibrium lobes are very different in nature. In particular, as shown in Fig. \ref{fig:dens}, the mean photon density is not constant within the lobes. Moreover, in the bistable region the two phases have very different photon density: one of them, hereafter called low-density phase, has  $\langle b^{\dagger} b\rangle\sim 10^{-2}$ (left panel of Fig. \ref{fig:dens}) whereas in the high density phase  $\langle b^{\dagger} b\rangle\gtrsim1$. (right panel of Fig. \ref{fig:dens}).

\begin{figure}[t!]
\includegraphics[width=0.49\columnwidth]{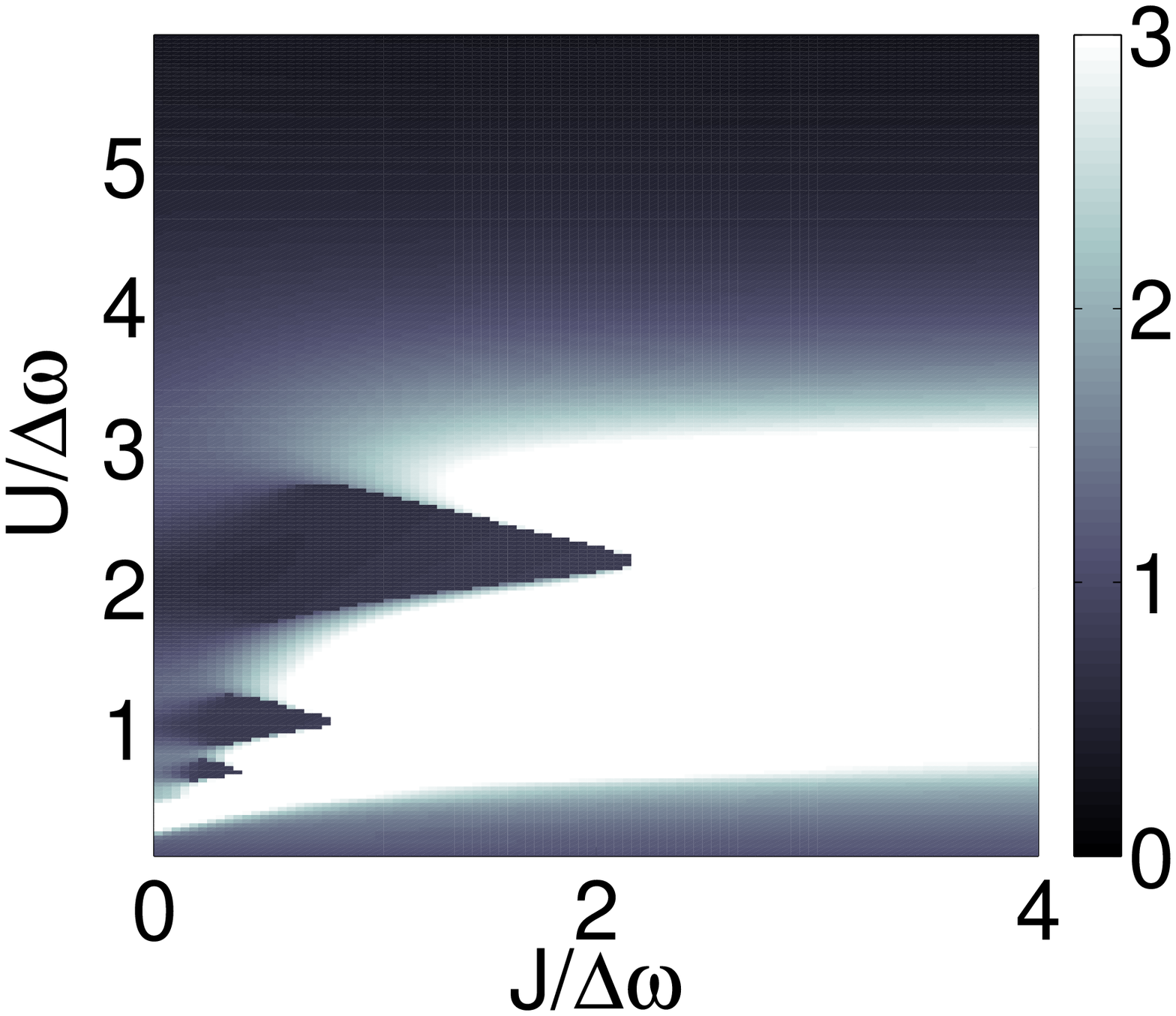}
\includegraphics[width=0.49\columnwidth]{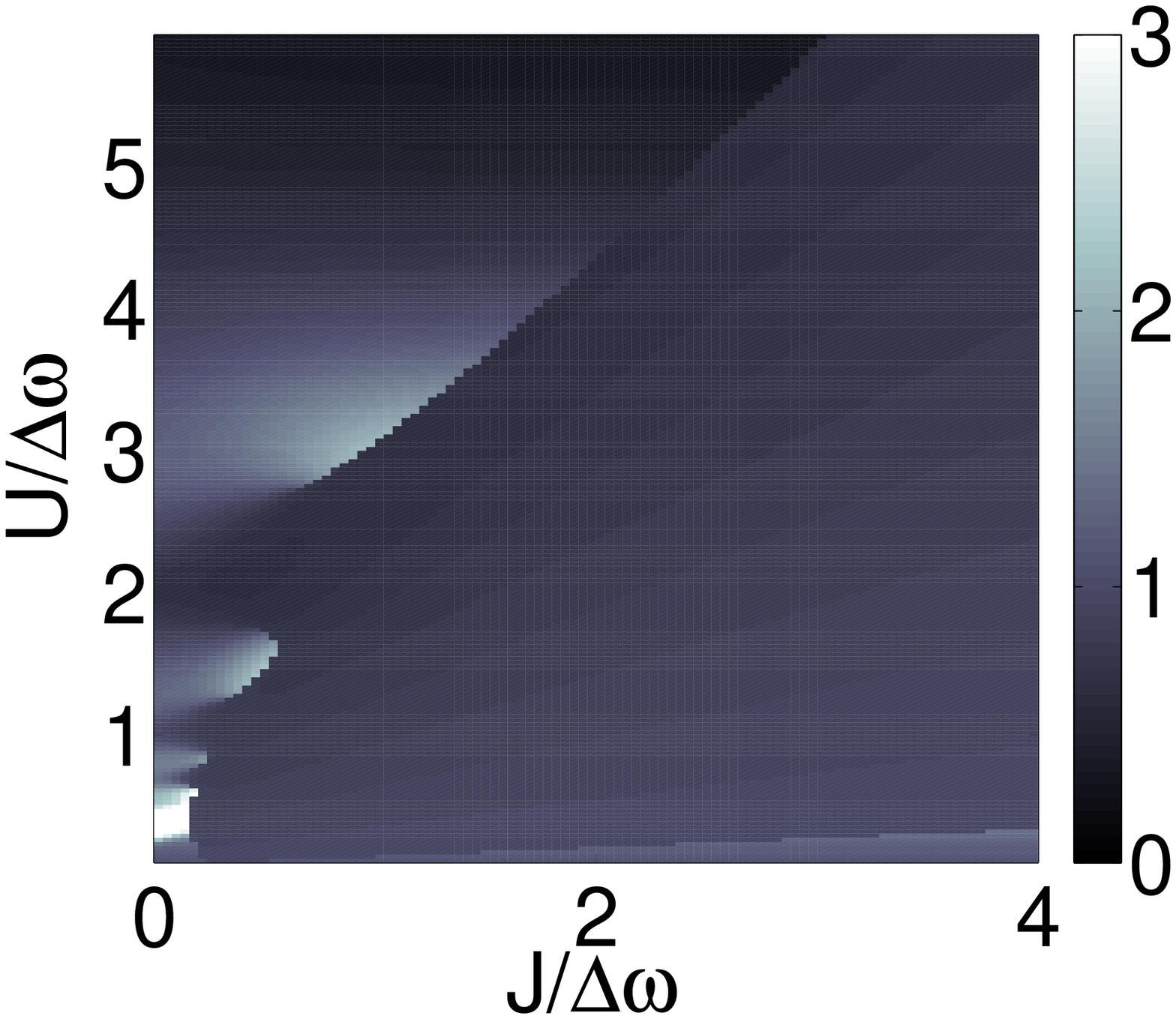}
\caption{ Second-order correlation function $g^2(0)$ as a function of $J/\Delta \omega$ and  $U/\Delta \omega$. Same parameters as in Fig. \ref{fig:dens}. Left panel for the low-density phase. Right panel for the high-density phase.}
\label{fig:g2}
\end{figure} 
To gain further insight, let us consider quantum correlations of the considered phases. The on-site second order correlation function $g^2(0)=\langle b^\dagger b^\dagger b b\rangle/\langle b^\dagger b\rangle ^2$  is plotted in Fig. \ref{fig:g2}. It is important to keep in mind that at equilibrium the value of $g^2(0)$ inside the lobes is equal to $1-1/n$ for a pure Mott insulator state where $n$ is the constant integer number of particles on each site, whereas it goes to 1 when $J \gg U$ in the thermodynamical limit. 
The left panel  of Fig. \ref{fig:g2} shows that the light is antibunched inside the lobes.  The lowest value, 0.6, is observed for the upper lobe, reminiscent of the $n=2$-lobe at equilibrium. But the analogy with equilibrium stops here : in the low density phase, for  $U/\Delta \omega < 3$ the emitted light shows strong bunching, that is $g^2(0) \gg 1$. Furthermore, there is no one-to-one correspondence between photon density and second order correlations. For example, in the low-density phase, for $U/\Delta \omega =  2$ and $J/\Delta \omega = 3.25$ the density is $0.026$ and $g^2(0) = 13$. For  $U/\Delta \omega =  4.5$ and $J/\Delta \omega =  1.6$, the density remains the same but $g^2(0) = 0.64$.
\begin{figure}[t!]
	\includegraphics[width = 0.49\columnwidth]{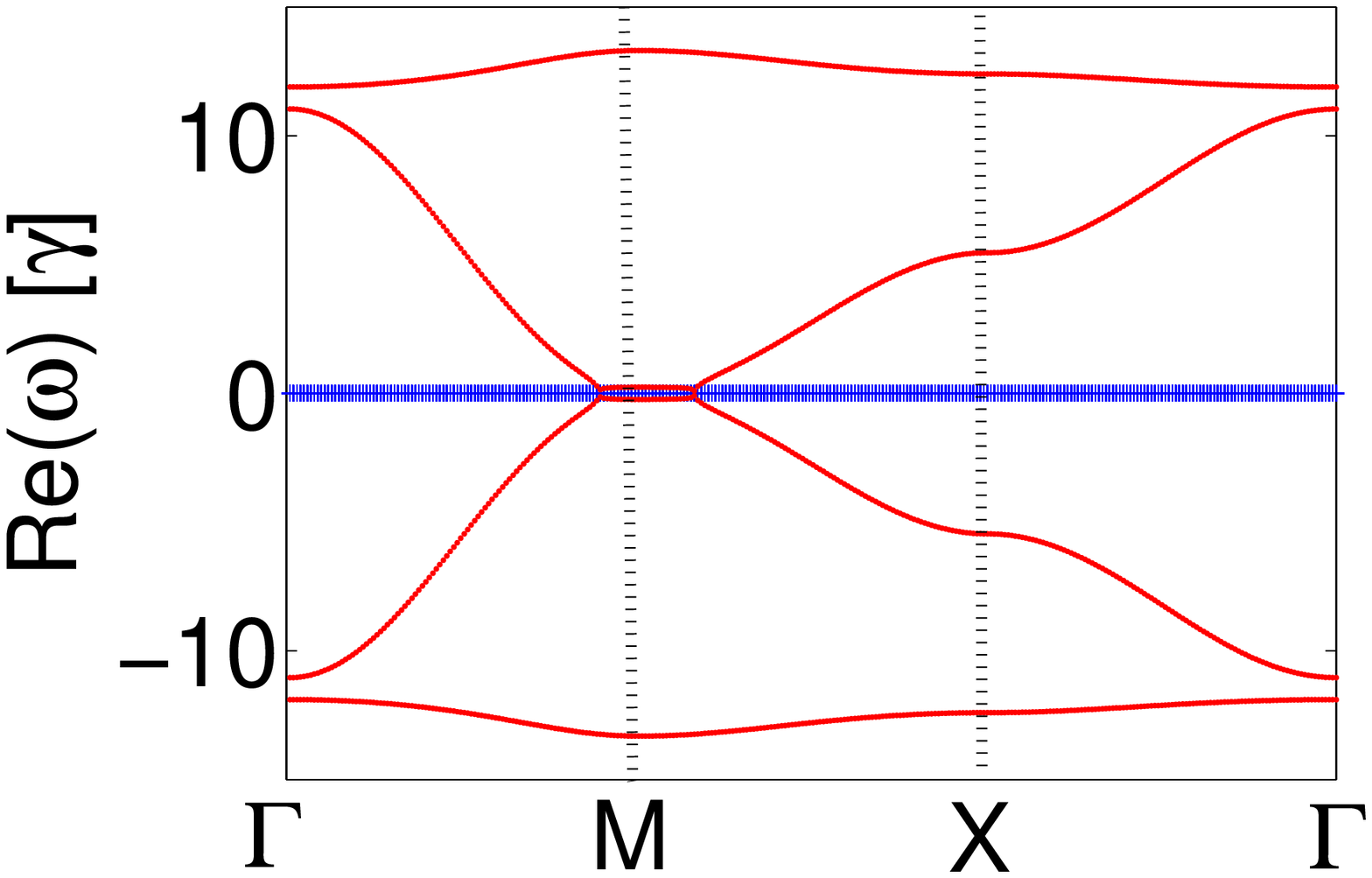}
	\includegraphics[width = 0.49\columnwidth]{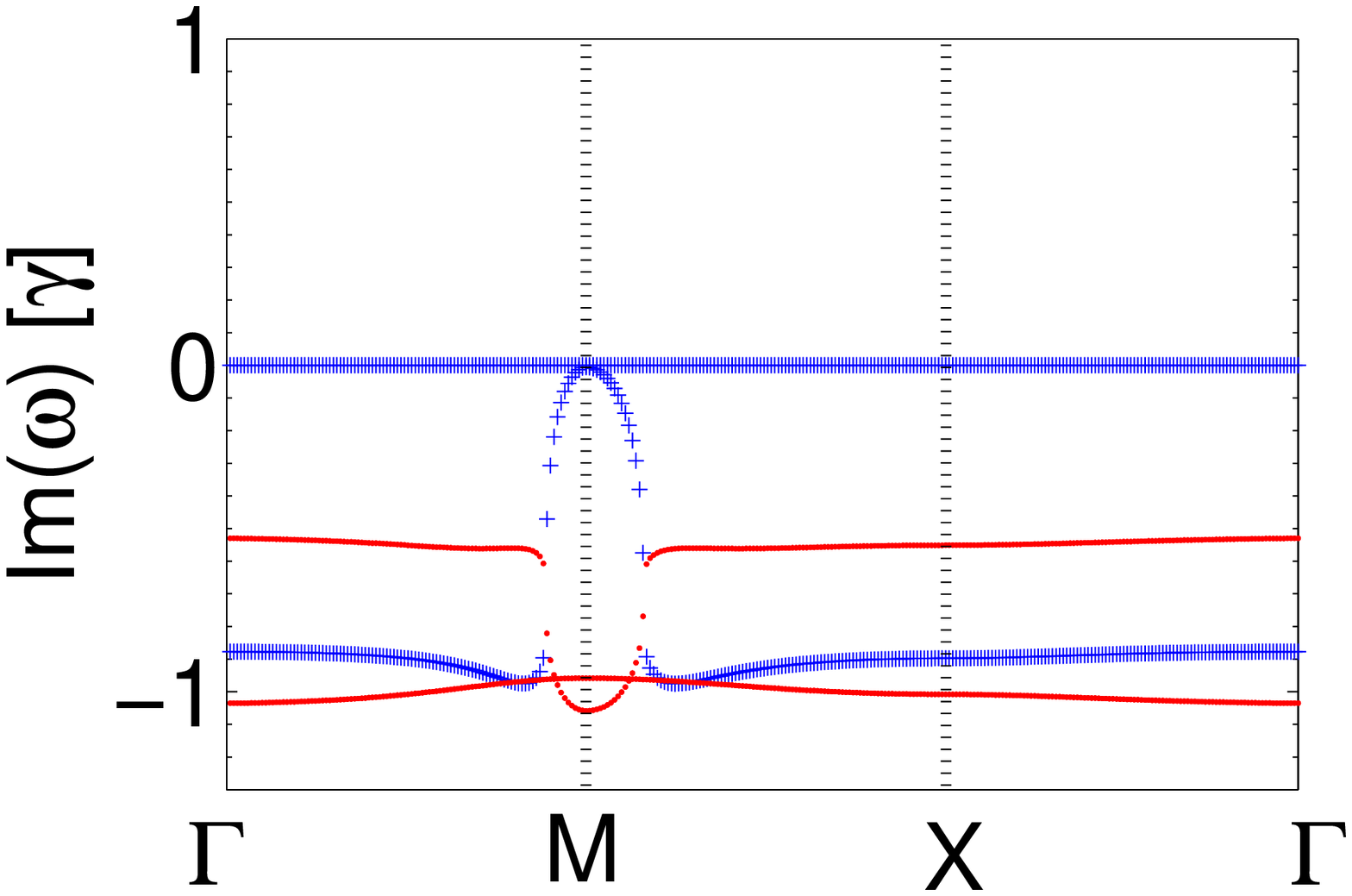}\\
	\includegraphics[width = 0.49\columnwidth]{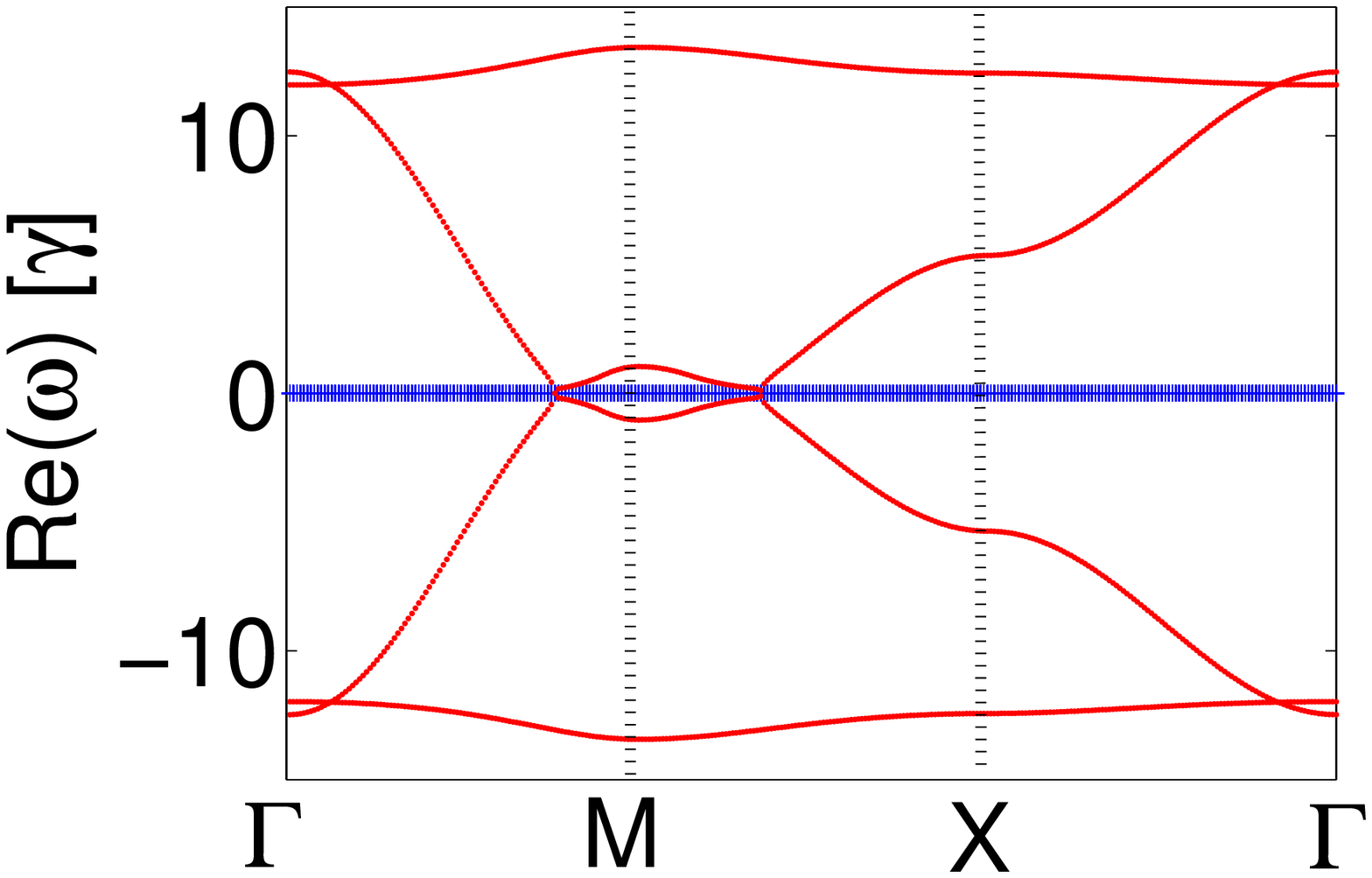}
	\includegraphics[width = 0.49\columnwidth]{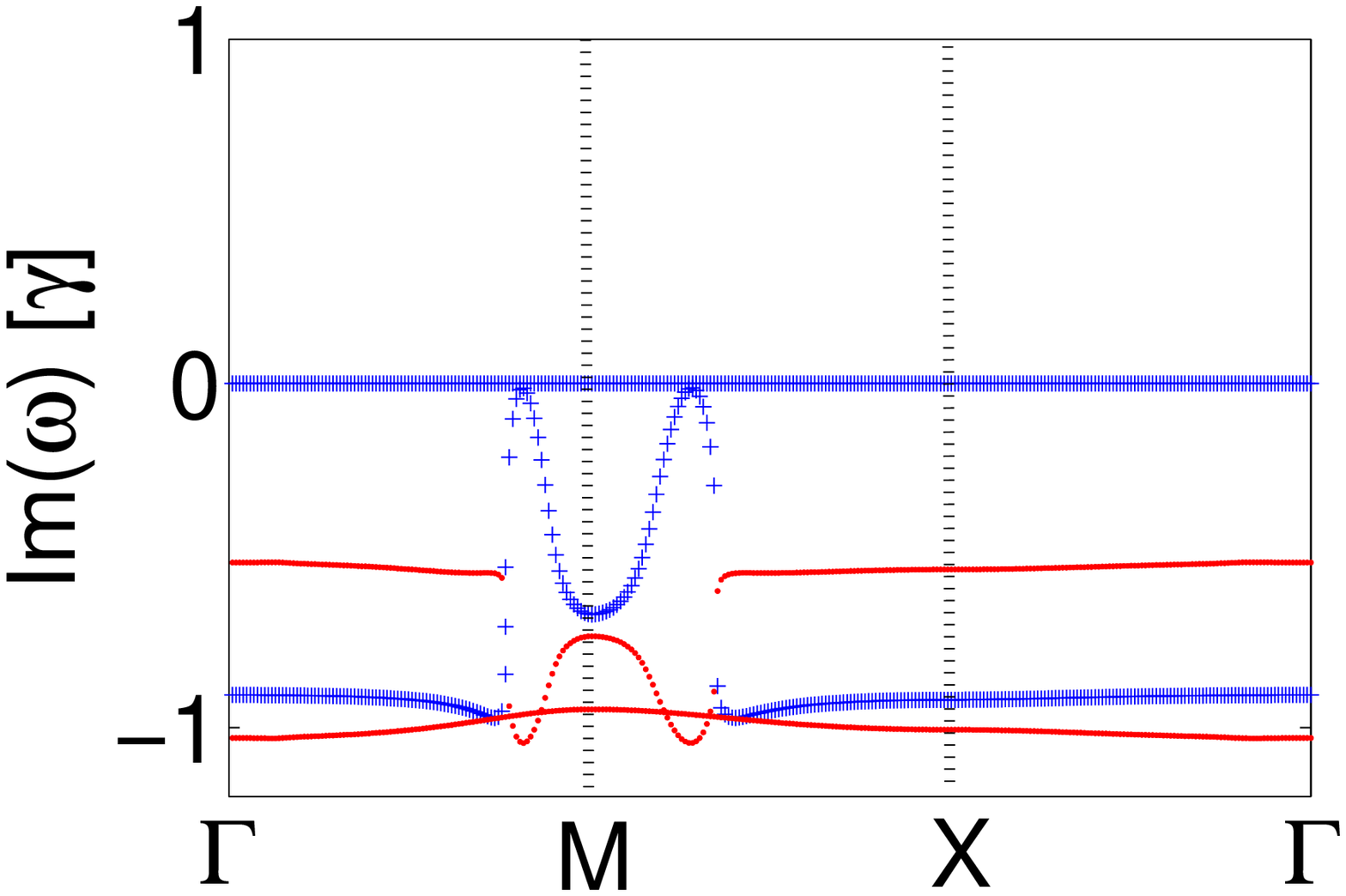}
\caption{Color online. Energy-momentum dispersion of elementary excitations for points A (upper panel) and A' (lower panel), indicated in Fig. \ref{fig:bistab}. Real and imaginary part of the low-energy branches (in units of $\gamma$)  are plotted vs $\mathbf{k}$. $\Gamma = (0,0)$, $M = (\pi/a, \pi/a)$, $X = (\pi/a,0)$ are special points in the Brillouin zone of the squared photonic lattice. Thick blue lines depict branches with a flat real part over the entire Brillouin zone, while the imaginary part is strongly dispersive with a resonance around specific wavevectors. }

\label{fig:bogoA}
\end{figure}     
\begin{figure}[t!]
	\includegraphics[width = 0.49\columnwidth]{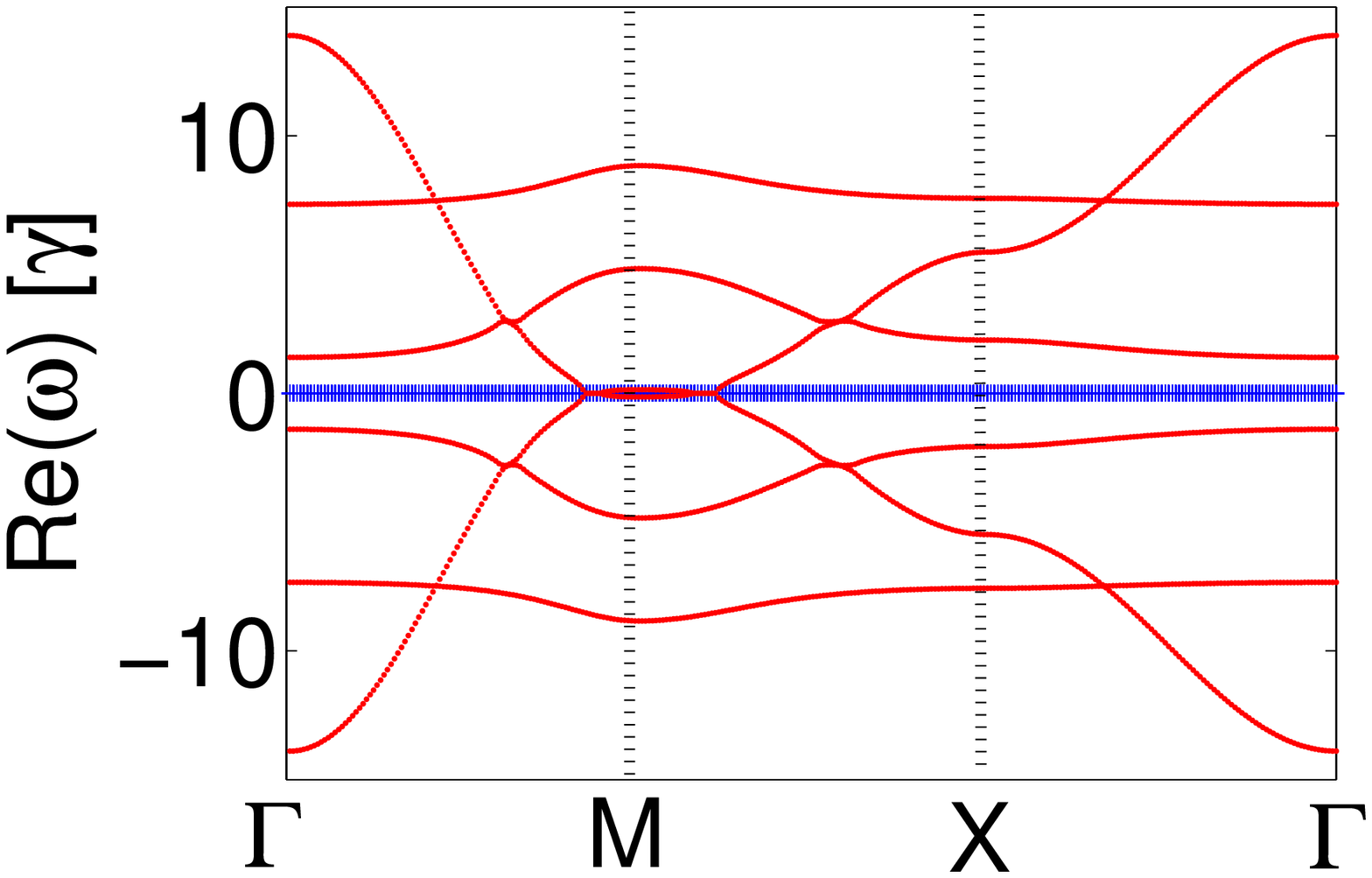}
	\includegraphics[width = 0.49\columnwidth]{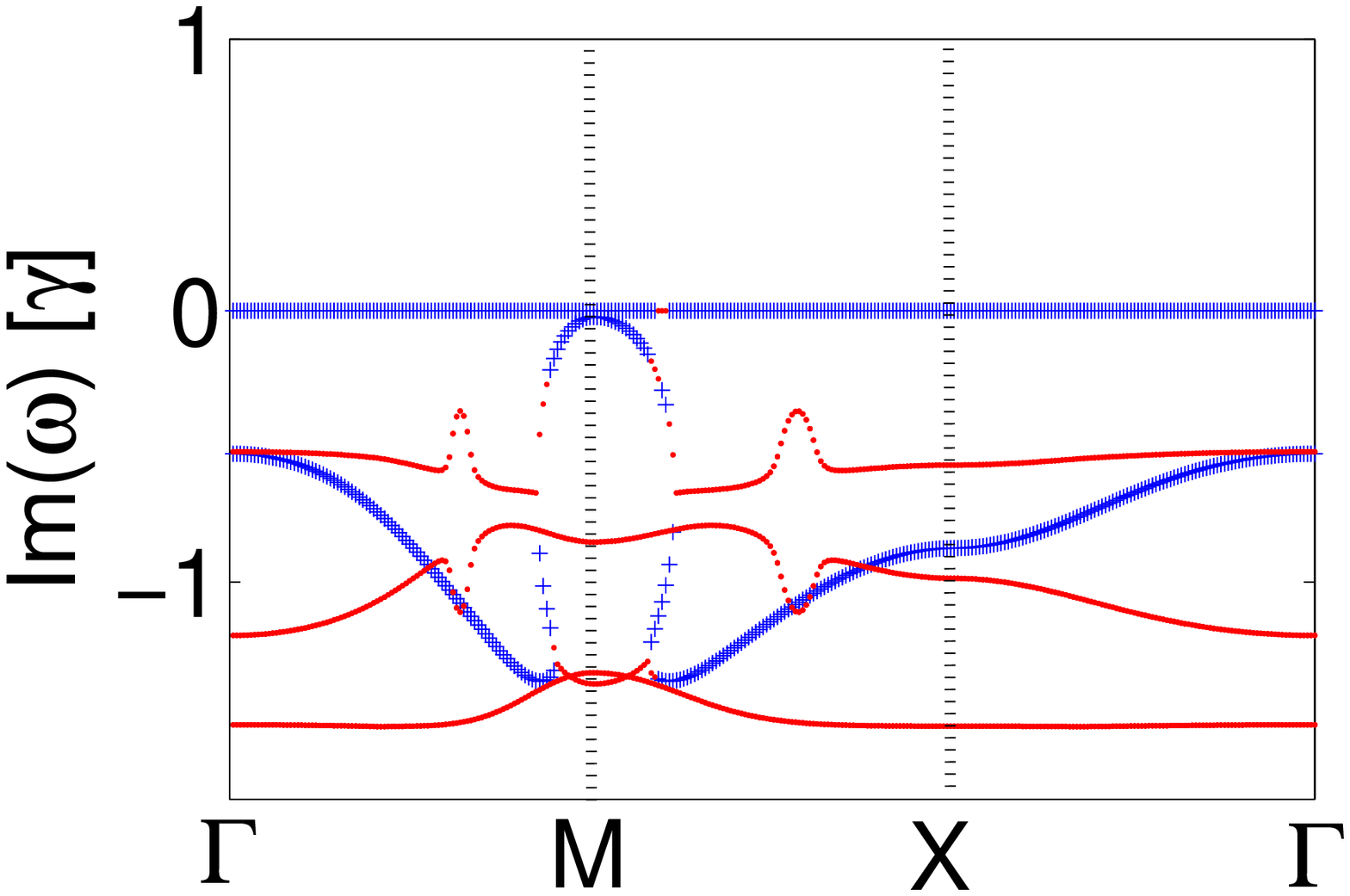}\\
	\includegraphics[width = 0.49\columnwidth]{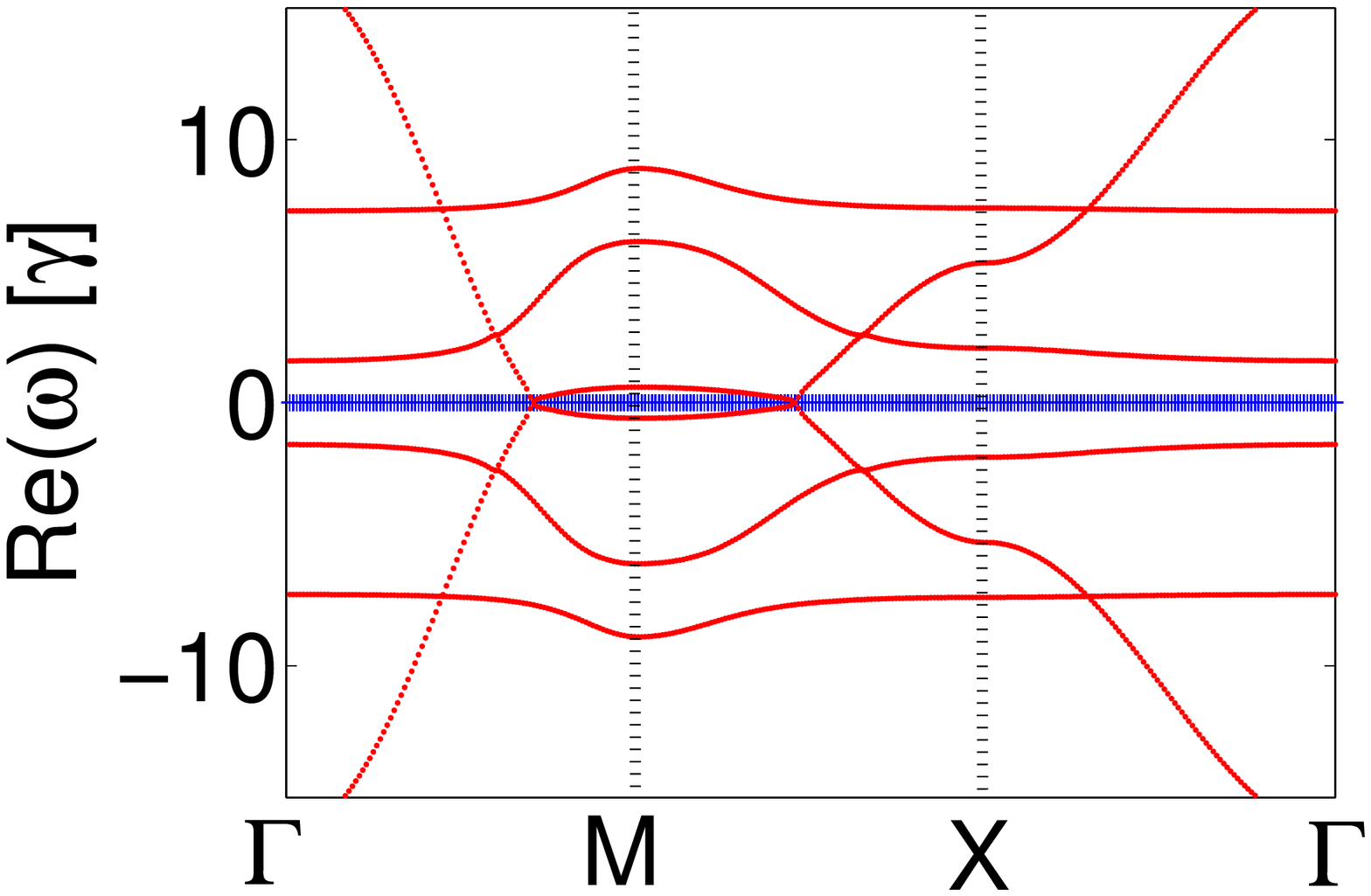}
	\includegraphics[width = 0.49\columnwidth]{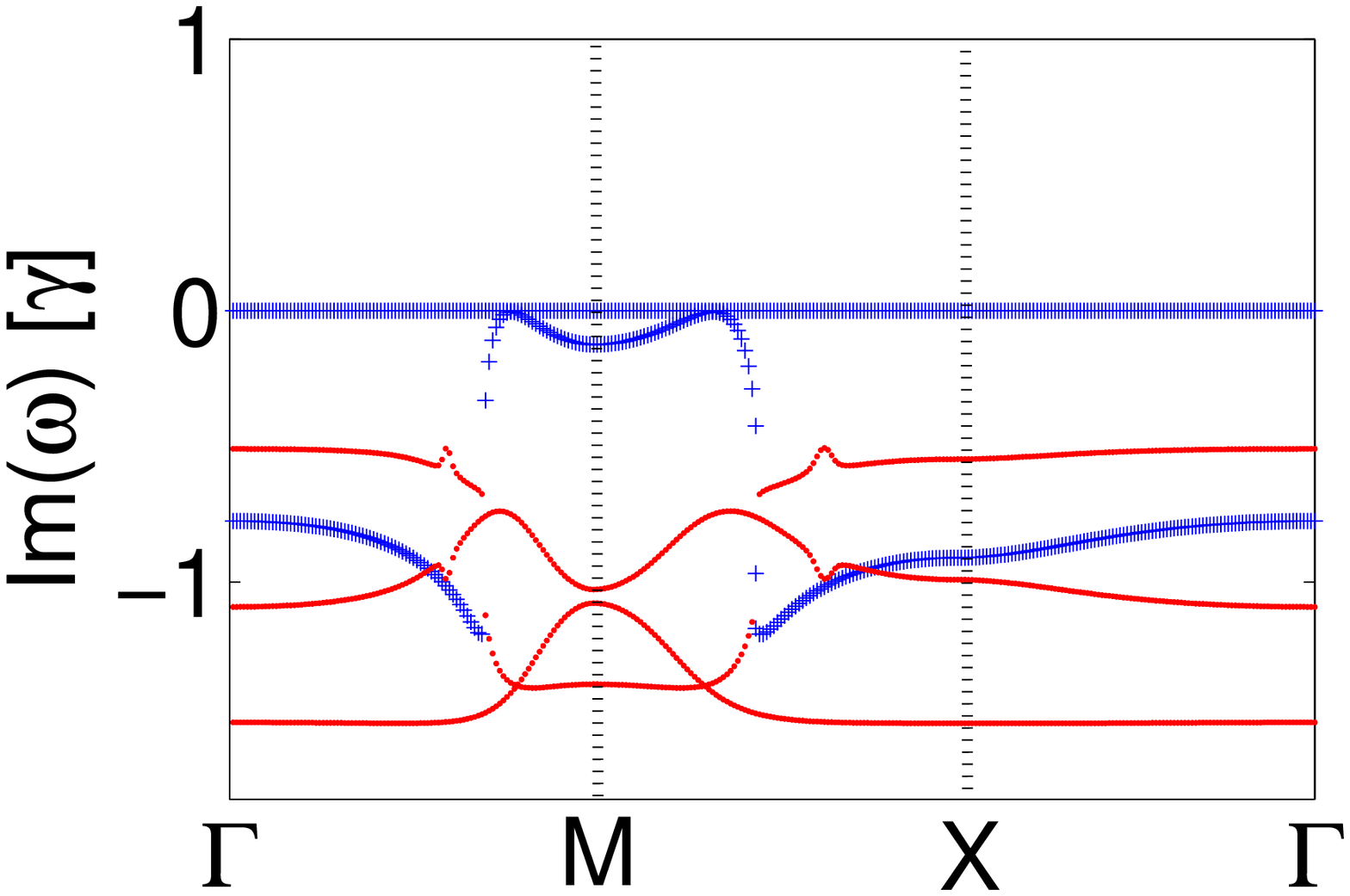}
\caption{Excitation dispersions for points B (upper panel) and B' (lower panel). Same conditions as in Fig. \ref{fig:bogoA}
\label{fig:bogoB}}
\end{figure}

Let us now investigate the collective excitations and the dynamical stability of the two phases. This can be done by linearizing the master equation in Fock space for small fluctuations around the steady-state  assuming a Gutzwiller factorization of the density matrix:
\begin{equation}\label{gutz}
\rho = \bigotimes_i (\overline{\rho} +\delta \rho^i),
\end{equation}
where $\overline \rho$ is the steady-state density matrix whose representation in Fock space can be extracted  
from Eq.(\ref{expect})\footnote{We have verified explicitly that $\overline \rho$ is indeed a steady-state solution of the master equation  (\ref{ME}) assuming the Gutzwiller ansatz (\ref{gutz})}. The coefficients of $\delta \rho^i$ then obey linear differential equations, which are coupled in real space but decoupled in reciprocal space due to the translational symmetry.  For each $\mathbf{k}$-vector in the Brillouin zone, we have:
\begin{equation}
i\partial_t   \delta \rho^{\mathbf{k}} = \mathcal{L}_{\mathbf{k}}.\delta \rho^{\mathbf{k}} ,\label{bogo}
\end{equation}
where $\delta \rho^{\mathbf{k}}_{n,m} = \frac{1}{\sqrt{N}}\sum_{i=1}^{N}e^{-i\mathbf{k\cdot r_i}}\delta \rho^i_{n,m}$ and $\mathcal{L}_{\mathbf{k}}$ is the matrix associated to the linearization. The energy spectrum is given by the eigenvalues of $\mathcal{L}_{\mathbf{k}}$ and the system is dynamically stable if all eigenvalues have negative imaginary part. 
 Stability studies revealed the onset of modulational instabilities in the low density phase (regions in yellow and red in Fig. \ref{fig:bistab}). Dispersion relations at the edge of the unstable region (points A, A', B and B' on Fig. \ref{fig:bistab}), are plotted on Fig. \ref{fig:bogoA} and \ref{fig:bogoB}. Remarkably, the real part of the unstable branch is zero for every $\mathbf{k}$ inside the Brillouin zone while the imaginary part is strongly dispersive. The existence of this purely imaginary branch can be seen analytically in the low-density regime
 where there is at most one photon per site so that  we can approximate our description by working in a truncated Hilbert space.  The vector $ \delta \rho^{\mathbf{k}}$ has then only four coefficients, $(\delta \rho^{\mathbf{k}}_{00}, \delta \rho^{\mathbf{k}}_{01},\delta \rho^{\mathbf{k}}_{11},\delta \rho^{\mathbf{k}}_{10})^T$ and $\mathcal{L}_{\mathbf{k}}$ is given by: 
 \begin{widetext} 
 \begin{equation}
 \mathcal{L_{\mathbf{k}}}  = 
 \begin{pmatrix}
 0 & ( A^{*}+F^{*})+ t_{\mathbf{k}}\overline{\rho}^*_{10}& i\gamma &-(A+F)- t_{\mathbf{k}}\overline{\rho}_{10}  \\ 
 A+F& - \Delta \omega - t_{\mathbf{k}}(\overline{\rho}_{00}-\overline{\rho}_{11})-i\frac{\gamma}{2} & -(A+F) &0 \\
0 & -( A^{*}+F^{*})- t_{\mathbf{k}}\overline{\rho}^*_{10} &-i\gamma& (A+F)+ t_{\mathbf{k}}\overline{\rho}_{10}  \\
 -(A^{*}+F^{*}) & 0 &A^{*}+F^{*} &\Delta \omega + t_{\mathbf{k}}(\overline{\rho}_{00}-\overline{\rho}_{11})-i\frac{\gamma}{2}
 \end{pmatrix}
 \end{equation}
 
 \end{widetext}
where $A = -J\langle b \rangle $ is the mean-field parameter, $(\overline{\rho}_{00},\overline{\rho}_{10},\overline{\rho}_{11})$  the coefficients of the steady-state density matrix, $t_{\mathbf{k}} =-J/2(\cos (k_x a )+ \cos(k_y a) )$
is the term responsible for the dispersion in a square lattice and $a$ is the lattice parameter. 

Simple algebra shows that such $4 \times 4$-matrix has always a purely imaginary eigenvalue. 
We have checked numerically that in the low density regime such truncated matrix agrees with the results obtained by including the full linearization matrix.
  The dispersive nature of purely imaginary branches is a consequence of interactions. Indeed, in the low-density phase, when $J \gg U$, all eigenvalues have completely flat imaginary parts, while there are only two (anti-conjugate) branches whose dispersive real parts originate from the bare boson dispersion on a square lattice.
The $\mathbf{k}$-dependence of $\mathcal{L}_{\mathbf{k}}$ is enclosed in the coefficient $t_{\mathbf{k}}$ which itself is a function of $\cos (k_x a)+ \cos (k_y a)$. We see on Fig. \ref{fig:bogoA} and Fig. \ref{fig:bogoB} that on the left side of the unstable region, instabilities arise at $\mathbf{k} = (\pi/a, \pi/a)$.  On the other edge  however, the unstable $\mathbf{k}$-vectors are smaller and located  well inside the Brillouin zone. The region marked in red on Fig. \ref{fig:bistab} is of particular interest as there is no stable homogenous solution inside. This shows that an inhomogeneous density-wave steady-state occurs in this region of the phase diagram.

In summary, we have explored the mean-field phase diagram of a driven-dissipative Bose-Hubbard model for a wide range of parameters. In the case of spatially homogenous coherent pumping, depending
on the values of the on-site repulsion and tunneling coupling, it is possible to have bistable or monostable homogeoneous solutions with peculiar quantum correlation properties. A collective excitation mode with 
a flat dispersion over the entire Brillouin zone and a dispersive imaginary part can occur, leading to tunneling-induced instabilities at specific wavevectors and thus a breaking of the translational invariance.
Our results shows that driven-dissipative arrays of cavities can lead to very rich manybody physics, which is very different from its equilibrium counterpart. 
C.C. is member of Institut Universitaire de France.

\end{document}